\documentclass[aps,prd,twocolumn,showpacs,groupedaddress,floatfix]{revtex4-1}

\usepackage{graphicx}
\usepackage{dcolumn}
\usepackage{bm}
\usepackage{amsmath}
\usepackage{amssymb}

\newcommand{\beq}{\begin{equation}}
\newcommand{\eeq}{\end{equation}}
\newcommand{\bea}{\begin{eqnarray}}
\newcommand{\eea}{\end{eqnarray}}

\begin{document}

\title{Chromoelectric flux tubes and coherence length in QCD}

\author{Paolo Cea}
\email{paolo.cea@ba.infn.it}
\affiliation{Dipartimento di Fisica dell'Universit\`a di Bari, I-70126 Bari, 
Italy \\
and INFN - Sezione di Bari, I-70126 Bari, Italy}

\author{Leonardo Cosmai}
\email{leonardo.cosmai@ba.infn.it}
\affiliation{INFN - Sezione di Bari, I-70126 Bari, Italy}

\author{Alessandro Papa}
\email{papa@cs.infn.it}
\affiliation{Dipartimento di Fisica dell'Universit\`a della Calabria,
I-87036 Arcavacata di Rende, Cosenza, Italy \\
and INFN - Gruppo collegato di Cosenza, I-87036 Arcavacata di Rende, Cosenza, 
Italy}

\date{\today}           

\begin{abstract}
The transverse profile of the chromoelectric flux tubes in SU(2) and SU(3) 
pure gauge theories is analyzed by a simple variational ansatz  using a 
strict analogy with ordinary superconductivity. Our method allows to extract the penetration length
and the coherence length of the flux tube.
\end{abstract}


\maketitle

\section{Introduction}

The presence of chromoelectric flux tubes in QCD vacuum is a clear signal of 
color confinement~\cite{Bander:1980mu,Greensite:2003bk}.
Monte Carlo simulations of lattice QCD can produce   a  sample of vacuum 
configurations, thus allowing a thorough nonperturbative study of tube-like 
structures that emerge by analyzing the chromoelectric fields between static 
quarks~\cite{Fukugita:1983du,Kiskis:1984ru,Flower:1985gs,Wosiek:1987kx,DiGiacomo:1989yp,DiGiacomo:1990hc,Singh:1993jj,Cea:1992sd,Matsubara:1993nq,Cea:1992vx,Cea:1993pi,Cea:1994ed,Cea:1994aj,Cea:1995zt,Bali:1994de,Haymaker:2005py,D'Alessandro:2006ug}.
A direct consequence of the tube-like structure of the chromoelectric fields 
between static quarks is the linear potential and hence the color confinement.

A striking physical analogy exists between the QCD vacuum and an electric 
superconductor. As conjectured long time ago by 't Hooft~\cite{'tHooft:1976ep} 
and Mandelstam~\cite{Mandelstam:1974pi}, the vacuum of QCD could be modeled as 
a coherent state of color magnetic monopoles, what is well known as dual 
superconductor~\cite{Ripka:2003vv}. In the dual superconductor model of QCD 
vacuum the condensation of color magnetic monopoles is analogous to the 
formation of Cooper pairs in the BCS theory of superconductivity. Even if the 
dynamical formation of color magnetic monopoles is not explained by the 
't Hooft construction, there is a lot of lattice evidences~\cite{Shiba:1994db,Arasaki:1996sm,Cea:2000zr,Cea:2001an,DiGiacomo:1999fa,DiGiacomo:1999fb,Carmona:2001ja,Cea:2004ux,D'Alessandro:2010xg} for the color magnetic condensation in 
QCD vacuum. It should be recognized~\cite{'tHooft:2004th} that the color 
magnetic monopole condensation in the confinement mode of QCD could be a 
consequence  rather  than the origin of the mechanism of color confinement,
that actually could be originated from additional dynamical causes. 
Notwithstanding the dual superconductivity picture of the QCD vacuum remains 
at least a very useful phenomenological frame to interpret the vacuum dynamics.

In the usual electric superconductivity tube-like structures 
  arise~\cite{Abrikosov:1957aa} as a solution of the Ginzburg-Landau 
equations. Similar solutions were found by Nielsen and 
Olesen~\cite{Nielsen:1973cs} in the case of the Abelian Higgs model, 
where they showed that a vortex solution exists independently of the type 
I or type II superconductor behavior of the vacuum.
In previous studies~\cite{Cea:1992vx,Cea:1993pi,Cea:1994ed,Cea:1994aj,Cea:1995zt,Cardaci:2010tb} performed by some of the present authors, color flux tubes 
made up of chromoelectric field directed along the line joining a static 
quark-antiquark pair has been investigated, in the cases of SU(2) and SU(3).

In the present work we would like to push forward the analogy with electric 
superconductivity and exploit some results~\cite{Clem:1975aa} in the 
superconductivity to further extract information from flux tube configurations 
in SU(2) and SU(3) vacuum. The method and the numerical results for both 
SU(2) and SU(3) are reported in Section~\ref{numerical}. In 
Section~\ref{lengths} we check the scaling of the penetration and coherence 
lengths for both SU(2) and SU(3), and  compare with previous studies. In 
Section~\ref{string-tension} we critically discuss the contribution of the 
longitudinal chromoelectric field to the string tension. Finally, in 
Section~\ref{conclusions} we summarize our results and present our 
conclusions. 

\section{Chromoelectric flux tubes on the lattice}
\label{numerical}

To explore on the lattice the field configurations produced by a static 
quark-antiquark pair we exploit the following connected correlation 
function~\cite{DiGiacomo:1989yp,DiGiacomo:1990hc,Kuzmenko:2000bq,DiGiacomo:2000va}
\begin{equation}
\label{rhoW}
\rho_W = \frac{\left\langle {\rm tr}
\left( W L U_P L^{\dagger} \right)  \right\rangle}
              { \left\langle {\rm tr} (W) \right\rangle }
 - \frac{1}{N} \,
\frac{\left\langle {\rm tr} (U_P) {\rm tr} (W)  \right\rangle}
              { \left\langle {\rm tr} (W) \right\rangle } \; ,
\end{equation}
where $U_P=U_{\mu\nu}(x)$ is the plaquette in the $(\mu,\nu)$ plane, connected
to the Wilson loop $W$ by a Schwinger line $L$, $N$ is the number of colors
(see Fig.~1 in Refs.~\cite{Cea:1995zt,Cardaci:2010tb}).
The correlation function defined in Eq.~(\ref{rhoW}) measures the field 
strength. Indeed, in the naive continuum limit~\cite{DiGiacomo:1990hc}
\begin{equation}
\label{rhoWlimcont}
\rho_W  \stackrel{a \rightarrow 0}{\longrightarrow} a^2 g \left[ \left\langle
F_{\mu\nu}\right\rangle_{q\bar{q}} - \left\langle F_{\mu\nu}
\right\rangle_0 \right]  \;,
\end{equation}
where $\langle\quad\rangle_{q \bar q}$ denotes the average in the presence of 
a static $q \bar q$ pair and $\langle\quad\rangle_0$  is the vacuum 
average. 
According to Eq.~(\ref{rhoWlimcont}), we define the color field strength
tensor as
\begin{equation}
\label{fieldstrength}
F_{\mu\nu}(x) = \sqrt\frac{\beta}{2 N} \, \rho_W(x)   \;.
\end{equation}
By varying the distance and the orientation of the plaquette $U_P$
with respect to the Wilson loop $W$, one can probe the color field
distribution of the flux tube. In particular, the case of plaquette
parallel to the Wilson loop corresponds to the component of the 
chromoelectric field longitudinal to the axis defined by the static quarks.
In previous studies the formation of chromoelectric flux tubes
was investigated in SU(2) lattice gauge theory~\cite{Cea:1992sd,Cea:1992vx,Cea:1993pi,Cea:1994ed,Cea:1994aj,Cea:1995zt} 
and in SU(3) lattice gauge theory~\cite{Cardaci:2010tb} by exploiting the 
connected correlation function Eq.~(\ref{rhoW}). It was found that the flux 
tube is almost completely formed by the longitudinal chromoelectric field, 
$E_l$, which is constant along the flux  axis  and decreases rapidly in 
the transverse direction $x_t$. By interpreting the formation of 
chromoelectric flux tubes as dual Meissner effect in the context of the dual 
superconductor model of confinement, the proposal was 
advanced~\cite{Cea:1992sd,Cea:1992vx,Cea:1993pi,Cea:1994ed,Cea:1994aj,Cea:1995zt} to fit the transverse shape of the longitudinal chromoelectric field 
according to
\begin{equation}
\label{London}
E_l(x_t) = \frac{\phi}{2 \pi} \mu^2 K_0(\mu x_t) \;,\;\;\;\;\; x_t > 0\;.
\end{equation}
Here, $K_0$ is the modified Bessel function of order zero, $\phi$ is
the external flux, and $\lambda=1/\mu$ is the London penetration length. 
Equation~(\ref{London}) is valid in the region $x_t \gg \xi$, $\xi$ being the 
coherence length which measures the coherence of the magnetic monopole 
condensate (the dual version of the Cooper condensate). In fact, we expect that
Eq.~(\ref{London}) gives an adequate description of the transverse structure 
of the flux tube if $\lambda \gg \xi$. This means that Eq.~(\ref{London}) 
should be valid for $\kappa \gg  1$ (type II superconductor), where $\kappa$ 
is the Ginzburg-Landau parameter,
\begin{equation}
\label{G-L-kappa}
\kappa = \frac{\lambda}{\xi} \; .
\end{equation}
However, several numerical studies~\cite{Suzuki:1988yq,Maedan:1989ju,Singh:1993ma,Singh:1993jj,Matsubara:1994nq,Schlichter:1997hw,Bali:1997cp,Schilling:1998gz,Gubarev:1999yp,Koma:2001ut,Koma:2003hv} in both SU(2) and SU(3) lattice 
gauge theories indicated that the vacuum behaves like an effective dual 
superconductor which belongs to the borderline between a type I and type II 
superconductor with $\kappa \sim 1$.
Thus, we see that Eq.~(\ref{London}) is no longer adequate to account for the 
transverse structure of the longitudinal chromoelectric field.
Remarkably, it turns out that we may re-analyze our lattice data for 
chromoelectric flux tubes by exploiting the results presented in
Ref.~\cite{Clem:1975aa} where, from the assumption of a simple variational 
model for the magnitude of the normalized order parameter of an isolated 
vortex, a simple analytic expression is derived  for the magnetic field 
and  supercurrent  density that solve Ampere's law and the Ginzburg-Landau 
equation.  In particular, the transverse distribution of the magnetic filed 
reduces to the London model results outside the vortex core, but has the added 
advantage of yielding  realistic values in the vortex core vicinity. 
Accordingly, from Eq.~(4) of Ref.~\cite{Clem:1975aa} we derive
\begin{equation}
\label{clem1}
E_l(x_t) = \frac{\phi}{2 \pi} \frac{1}{\lambda \xi_v} \frac{K_0(R/\lambda)}
{K_1(\xi_v/\lambda)} \;,
\end{equation}
with
\begin{equation}
\label{rrr}
 R=\sqrt{x_t^2+\xi_v^2}    \;,
\end{equation}
where $\xi_v$ is a variational core radius parameter found to 
be~\cite{Clem:1975aa} of the order of $\xi$. 
Equation~(\ref{clem1}) can be written as
\begin{equation}
\label{clem2}
E_l(x_t) =  \frac{\phi}{2 \pi} \frac{\mu^2}{\alpha} \frac{K_0[(\mu^2 x_t^2 
+ \alpha^2)^{1/2}]}{K_1[\alpha]} \; ,
\end{equation}
with
\begin{equation}
\label{alpha}
\mu= \frac{1}{\lambda} \,, \quad \frac{1}{\alpha} =  \frac{\lambda}{\xi_v} \,.
\end{equation}
By fitting Eq.~(\ref{clem2}) to our flux tubes data, we may obtain both the 
penetration length $\lambda$ and the ratio of the penetration length to the 
variational core radius parameter  $\lambda/\xi_v$. It is worth to recall that, by means of 
Eq.~(\ref{clem2}), we can extend our fit up to $x_t=0$.
Moreover by using Eq.~(16) of Ref.~\cite{Clem:1975aa}, we may also obtain the 
Ginzburg-Landau $\kappa$ parameter,
\begin{equation}
\label{landaukappa}
\kappa = \frac{\sqrt{2}}{\alpha} \left[ 1 - K_0^2(\alpha) / K_1^2(\alpha) 
\right]^{1/2} \,,
\end{equation}
with $K_1$ the modified Bessel function of order 1. 
The coherence length $\xi$ is obtained from Eqs.~(\ref{G-L-kappa}) and (\ref{landaukappa}).
Our data for chromoelectric fields between static quark-antiquark sources have 
been obtained through the connected correlation function Eq.~(\ref{rhoW}). 
In order to reduce the quantum fluctuations we adopted the controlled
cooling algorithm. It is known~\cite{Campostrini:1989ts} that 
by cooling in a smooth way equilibrium configurations, quantum fluctuations 
are reduced by a few order of magnitude, while the string tension survives 
and shows a plateau. We shall show below that the penetration length behaves 
in a similar way. The details of the cooling procedure are described in 
Ref.~\cite{Cea:1995zt} for the case of SU(2). Here we adapted the procedure
to the case of SU(3), by applying successively this algorithm to various 
SU(2) subgroups. The control parameter $\delta$ was fixed at the value 
0.0354, as in Ref.~\cite{Cea:1995zt}. As described in 
Ref.~\cite{Cardaci:2010tb}, in the construction of the lattice operator given 
in Eq.~(\ref{rhoW}) we have considered also noninteger distances to 
check the restoration of the rotational symmetry on our lattices.

\subsection{SU(2) data}
\label{su2data}

We analyzed our lattice SU(2) data collected for three different values of 
$\beta$, namely $\beta=2.52,2.55,2.6$  (for further details we refer to  
Ref.~\cite{Cardaci:2010tb}). Indeed, we find that Eq.~(\ref{clem2}) is able to 
reproduce the transverse distribution of the longitudinal chromoelectric field 
in the whole region $x_t \ge 0$.
\begin{figure}[htb]
\includegraphics*[width=0.95\columnwidth,clip]{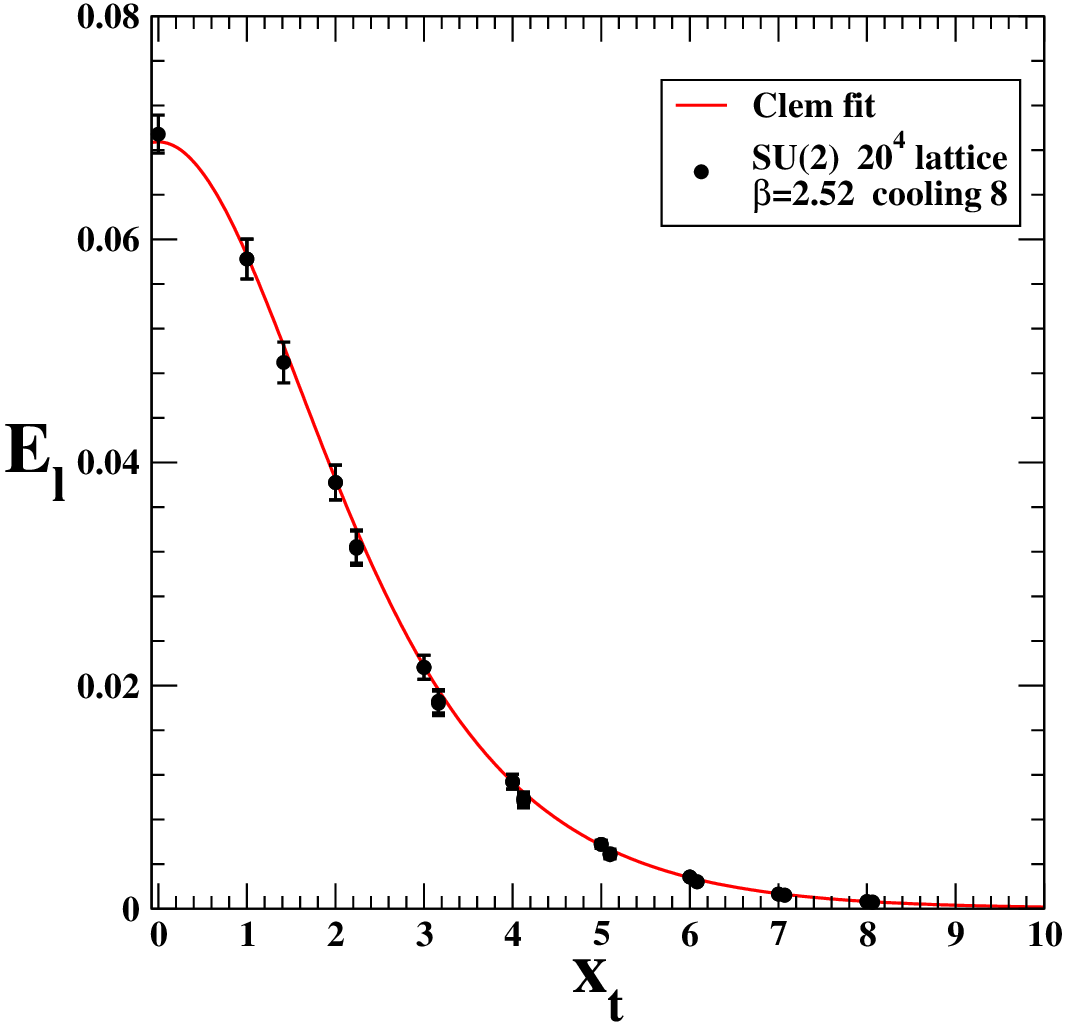} 
\caption{SU(2): $E_l$ versus $x_t$ for $\beta=2.52$ after 8 cooling steps.}
\label{El_SU2}
\end{figure}
An example of the of effectiveness of Eq.~(\ref{clem2}) to fit all data for 
the transverse distribution of the chromoelectric field down to $x_t=0$ is 
given in Fig.~\ref{El_SU2}, where we also display the points calculated at 
noninteger  distances, which were not included in the fit.
We see that there are slight deviations from the 
fit curve due to the failure of rotational invariance on a discrete lattice.  
In fact, fitting all the available data to Eq.~(\ref{clem2}) results in an 
increase of the reduced chi-squared without affecting appreciably the fit 
parameters.
\begin{table}[htb]
\begin{center}
\begin{tabular}{|c|c|c|c|c|c|}
\hline
\multicolumn{6}{|c|}{SU(2)\ \ \ $\beta=2.52$} \\
\hline
cooling & $\phi$ & $\mu$ & $\lambda/\xi_v$ & $\kappa$ & $\chi^2_r$  \\
\hline
  5 &   0.886( 89) &   0.829(267) &   0.590(452) &   0.512(400) &    0.3 \\
  6 &   1.214( 66) &   0.782(145) &   0.567(270) &   0.485(365) &    0.2 \\
  7 &   1.590( 53) &   0.749( 82) &   0.538(162) &   0.453(325) &    0.2 \\
  8 &   1.998( 49) &   0.711( 51) &   0.530(110) &   0.444(314) &    0.1 \\
  9 &   2.423( 49) &   0.685( 39) &   0.508( 82) &   0.420(285) &    0.1 \\
 10 &   2.680( 23) &   0.652( 22) &   0.530( 44) &   0.444(313) &    0.3 \\
\hline
\end{tabular}
\end{center}
\caption{Summary of the fit values for SU(2) at $\beta=2.52$.} 
\label{Table:2.52}
\end{table}
In Table I the results of our fit of Eq.~(\ref{clem2}) to the SU(2) data 
at $\beta=2.52$ are reported. The Ginzburg-Landau parameter $\kappa$ has been
obtained through Eq.~(\ref{landaukappa}).
\begin{figure}[htb]
\includegraphics*[width=0.95\columnwidth,clip]
{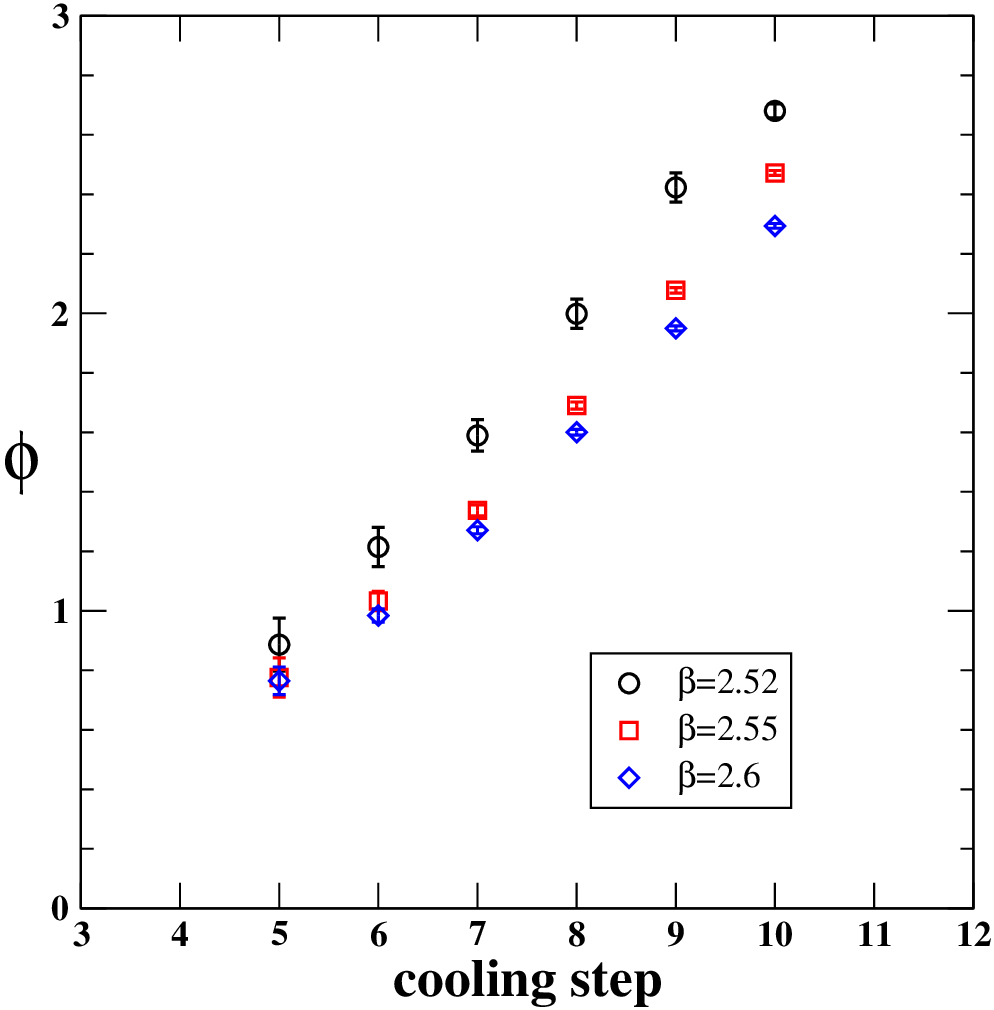} 
\caption{SU(2): $\phi$ versus cooling.}
\label{phiSU2}
\end{figure}
\begin{figure}[htb]
\includegraphics*[width=0.95\columnwidth,clip]
{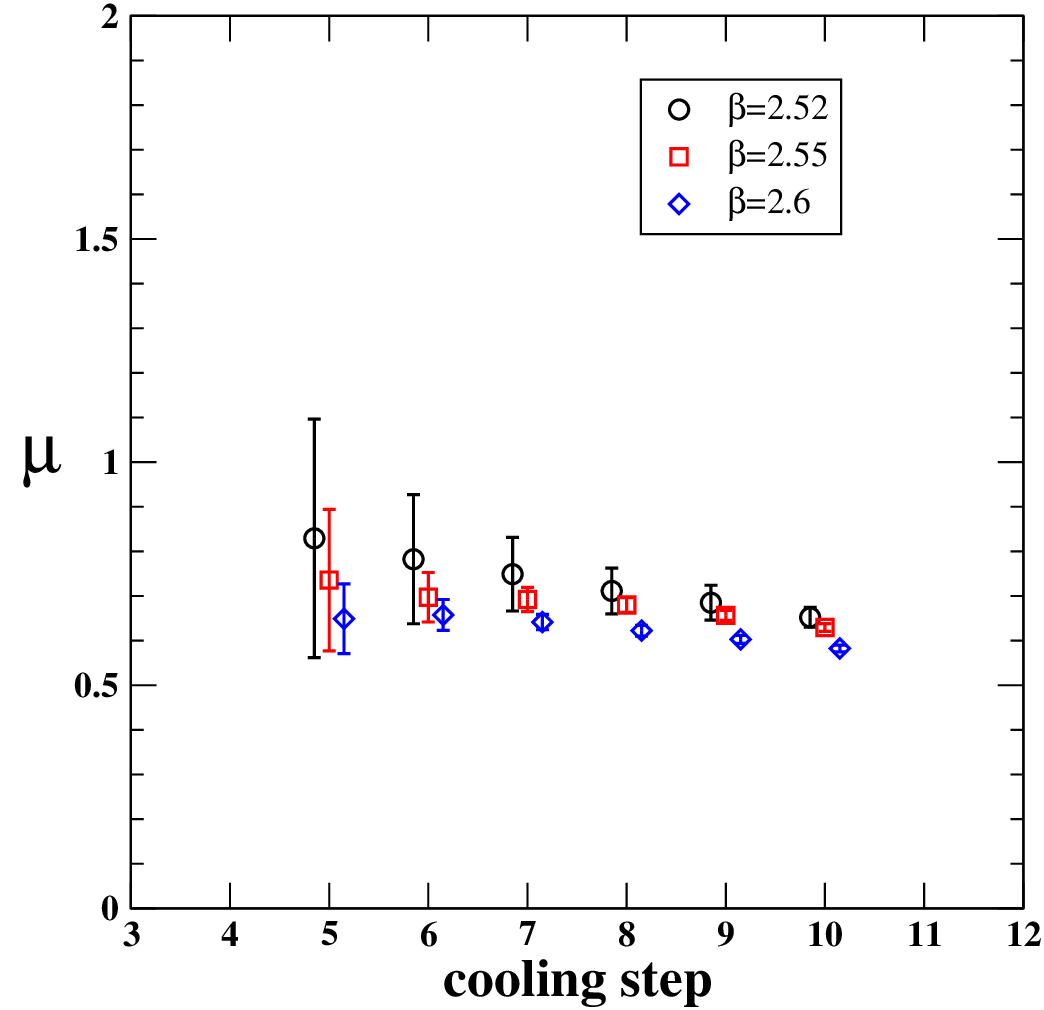} 
\caption{SU(2): $\mu$ versus cooling. Here and in other figures below
data have been slightly shifted along the horizontal axis for the sake of 
readability.}
\label{muSU2}
\end{figure}
\begin{figure}[htb]
\includegraphics*[width=0.95\columnwidth,clip]
{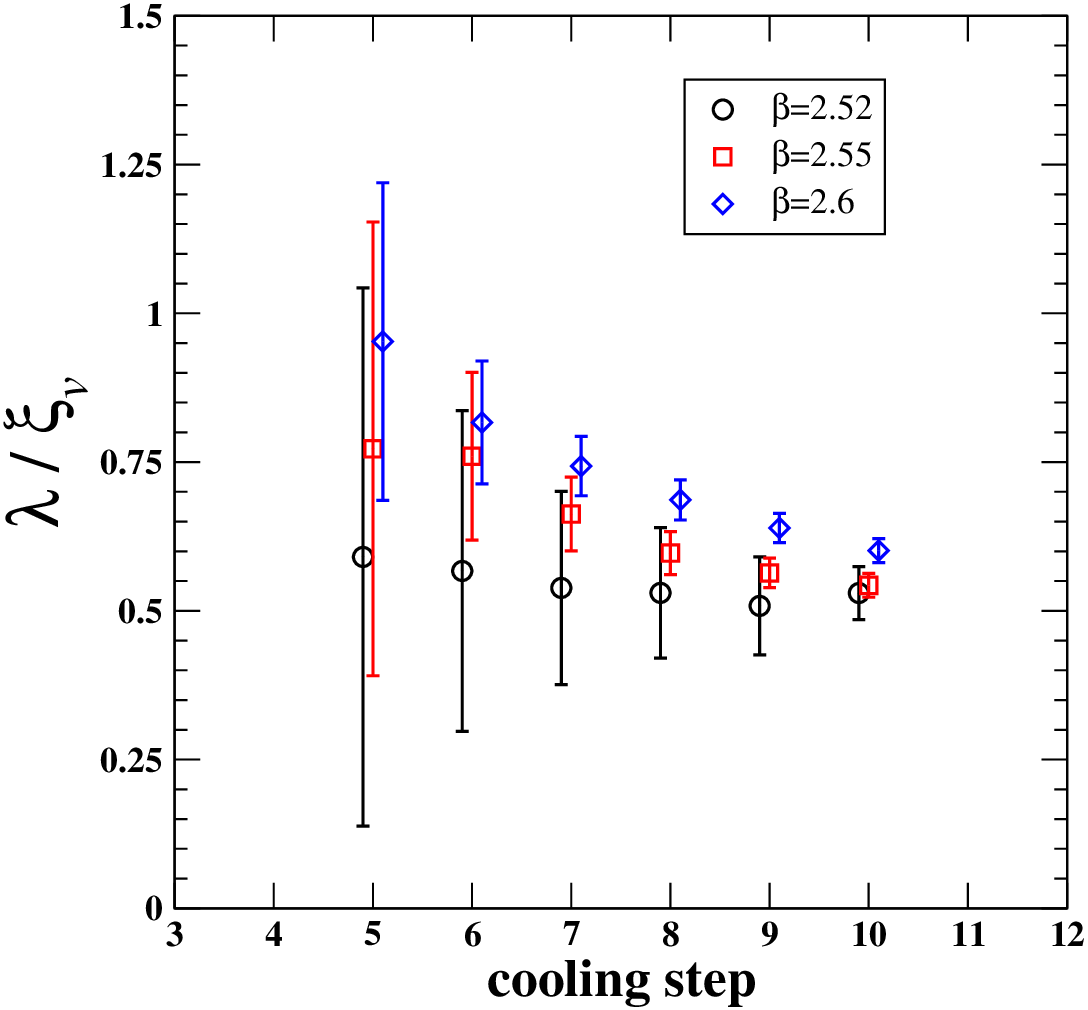} 
\caption{SU(2): $\lambda/\xi_v$ versus cooling.}
\label{lambdasucsivSU2}
\end{figure}
\begin{figure}[htb]
\includegraphics*[width=0.95\columnwidth,clip]
{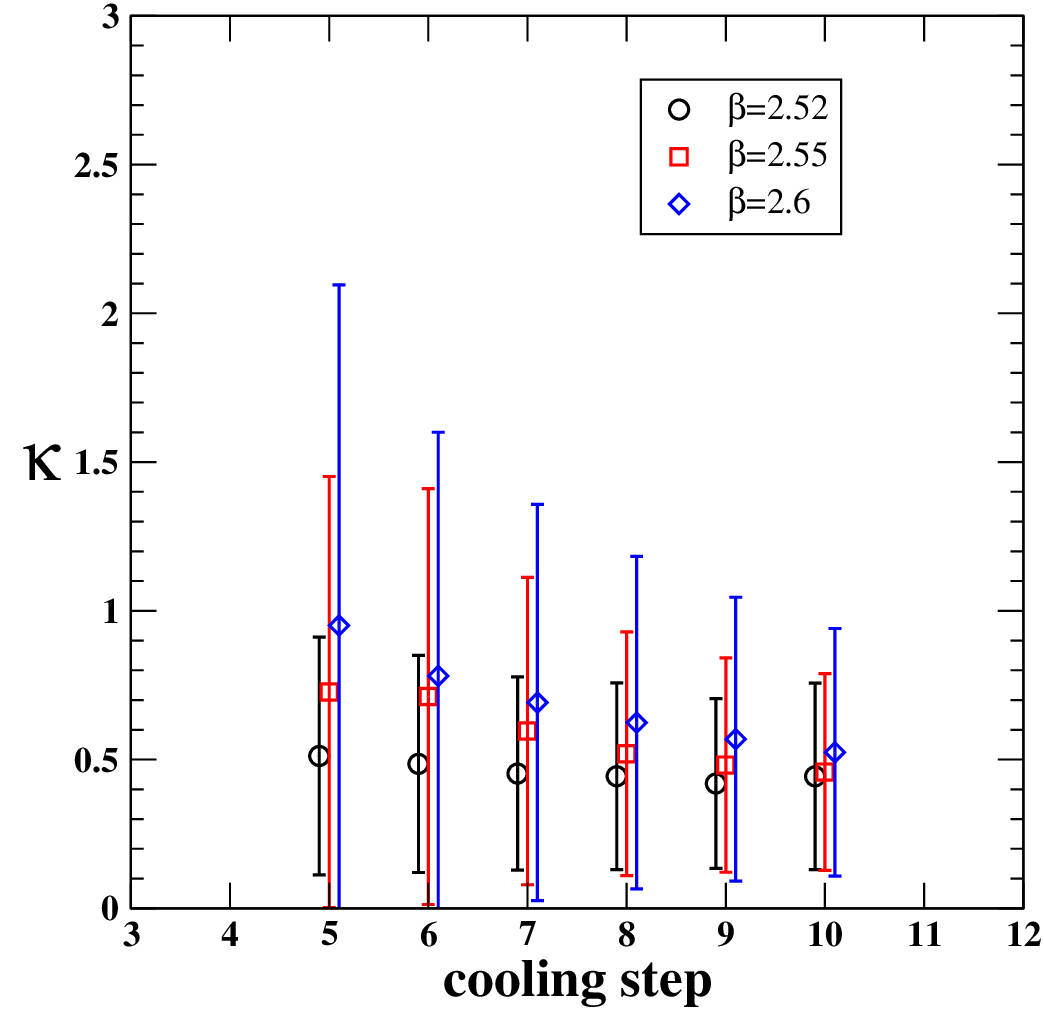} 
\caption{SU(2): $\kappa$ versus cooling.}
\label{kappaSU2}
\end{figure}
\\
In Figs.~\ref{phiSU2},  \ref{muSU2},  \ref{lambdasucsivSU2},  \ref{kappaSU2} 
we display the fitted parameters versus the cooling steps.  
As regards the parameters $\mu$, $\lambda/\xi_v$, and  $\kappa$, a short 
plateau is always visible. This corroborates our expectation that the long 
range physics is unaffected by the cooling procedure. On the other hand, 
Fig.~\ref{phiSU2} shows that the overall normalization of the transverse 
distribution of the longitudinal chromoelectric field is more affected by the 
cooling. In fact, the parameter  $\phi$ seems to displays an approximate 
short plateau after 9 - 10 cooling steps in accordance with previous 
studies~\cite{Cea:1995zt}.
\subsection{SU(3) data}
\label{su3data}
\begin{figure}[htb]
\includegraphics*[width=0.95\columnwidth,clip]
{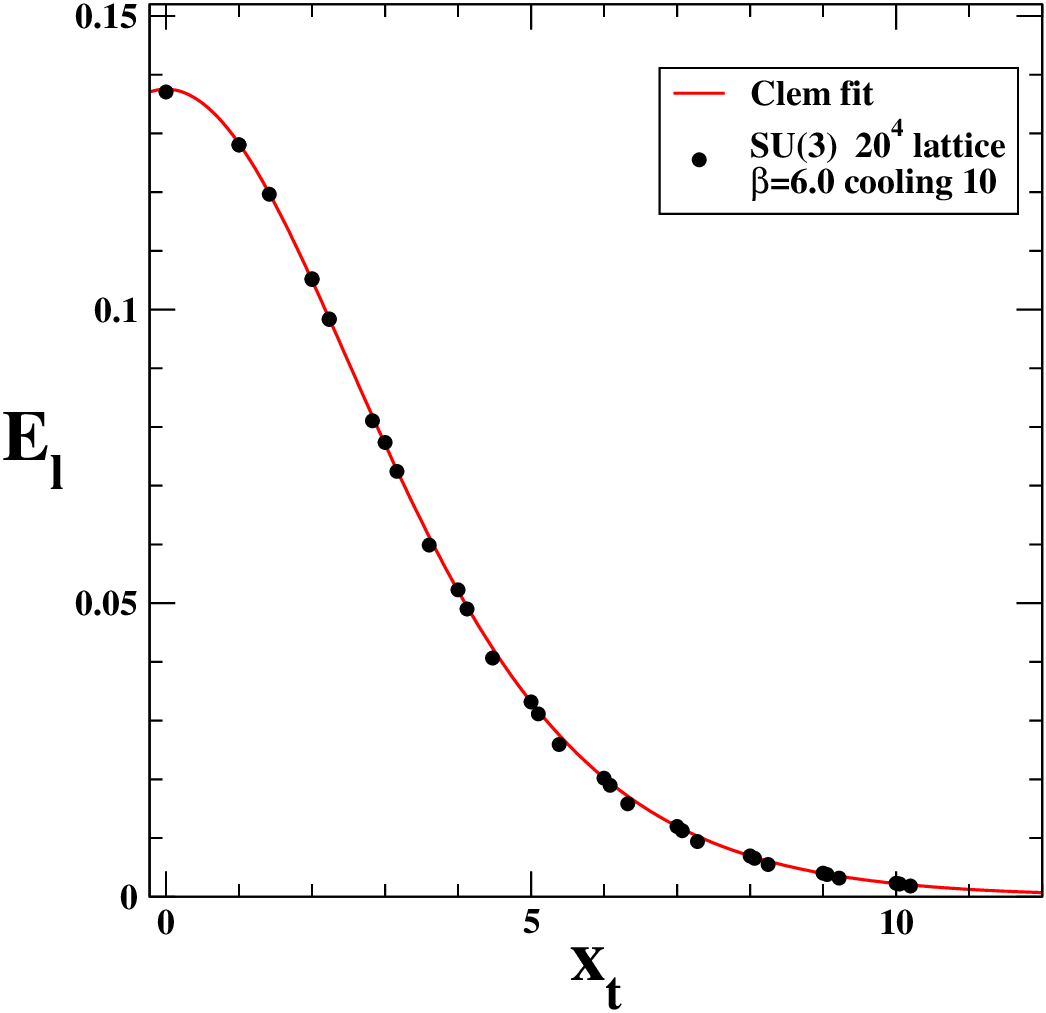} 
\caption{SU(3): $E_l$ at $\beta=6.0$ after 10 cooling steps.}
\label{El_SU3}
\end{figure}
We re-analyzed the SU(3) lattice data presented in 
Ref.~\cite{Cardaci:2010tb}. We have fitted the longitudinal chromoelectric 
field transverse distribution to Eq.~(\ref{clem2}) for 
$\beta=5.9,6.0,6.05,6.1$ and up to 16 cooling steps. Again, we find that  
Eq.~(\ref{clem2}) accounts for the transverse distribution of the 
longitudinal chromoelectric field in the whole region $x_t \ge 0$.  
In Fig.~\ref{El_SU3} we display our data for the transverse shape of the
longitudinal chromoelectric field between static quark-antiquark sources
after 10 cooling steps at $\beta=6.0$ together with the fit to 
Eq.~(\ref{clem2}).  As for the SU(2) case we also display the points 
calculated at noninteger distances and checked that the fit to all 
the available data of  Eq.~(\ref{clem2}) does not change the values of 
the fit parameters.
\begin{table}[htb]
\begin{center}
\begin{tabular}{|c|c|c|c|c|c|}
\hline
\multicolumn{6}{|c|}{SU(3) \ \ \ $\beta=6.0$} \\
\hline
cooling & $\phi$ & $\mu$ & $\lambda/\xi_v$ & $\kappa$ & $\chi^2_r$ \\
\hline
  5 &   4.564( 14) &   0.726(  8) &   0.516( 14) &   0.428(295) &    2.8 \\
  6 &   6.617(  9) &   0.683(  4) &   0.449(  7) &   0.355(216) &    0.3 \\
  7 &   7.644(  9) &   0.652(  4) &   0.412(  6) &   0.316(177) &    2.2 \\
  8 &   8.107(  8) &   0.630(  3) &   0.384(  5) &   0.287(150) &    4.6 \\
  9 &   8.254(  8) &   0.612(  3) &   0.360(  4) &   0.263(130) &    5.5 \\
 10 &   8.227(  7) &   0.598(  3) &   0.341(  4) &   0.244(114) &    5.2 \\
 11 &   8.108(  7) &   0.585(  3) &   0.325(  4) &   0.229(103) &    4.5 \\
 12 &   7.943(  7) &   0.574(  3) &   0.313(  4) &   0.217( 94) &    3.6 \\
 13 &   7.759(  6) &   0.563(  3) &   0.304(  4) &   0.208( 88) &    2.8 \\
 14 &   7.570(  6) &   0.553(  3) &   0.296(  4) &   0.201( 83) &    2.1 \\
 15 &   7.383(  6) &   0.544(  3) &   0.291(  4) &   0.196( 79) &    1.5 \\
 16 &   7.204(  6) &   0.534(  3) &   0.287(  4) &   0.192( 77) &    1.1 \\
\hline
\end{tabular}
\end{center}
\caption{Summary of the fit values for SU(3) at $\beta=6.0$.} 
\label{Table:6.0}
\end{table}
\begin{figure}[htb]
\includegraphics*[width=0.95\columnwidth,clip]
{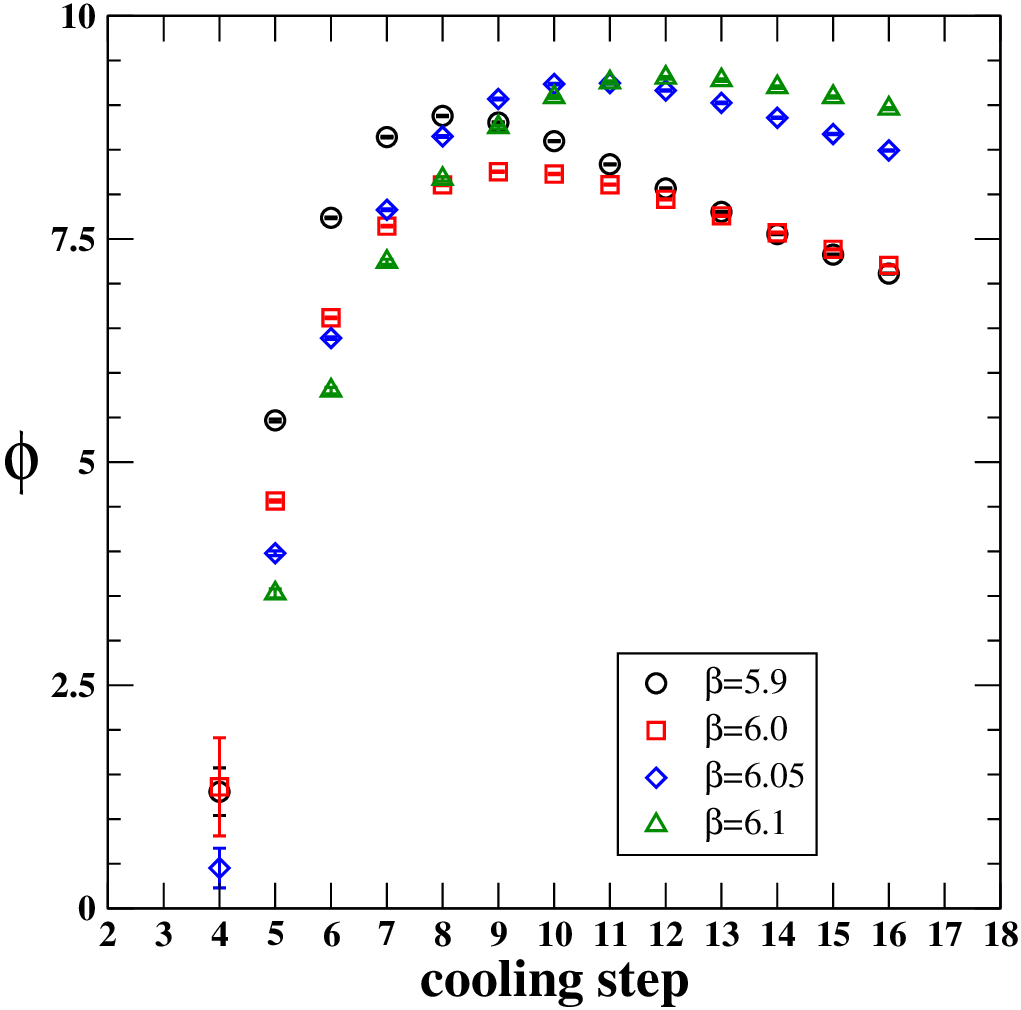} 
\caption{SU(3): $\phi$ versus cooling.}
\label{phiSU3}
\end{figure}

\begin{figure}[htb]
\includegraphics*[width=0.95\columnwidth,clip]
{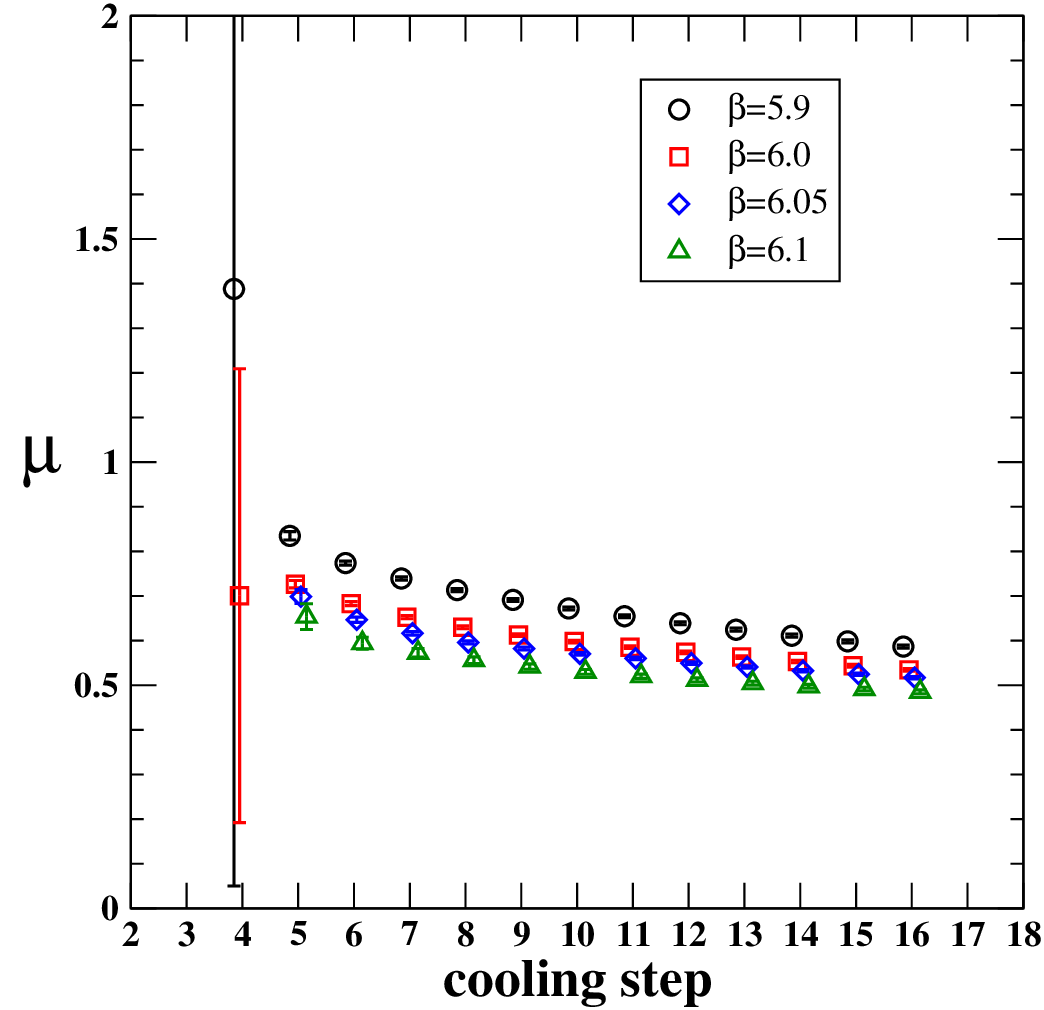} 
\caption{SU(3): $\mu$ versus cooling.}
\label{muSU3}
\end{figure}

\begin{figure}[htb]
\includegraphics*[width=0.95\columnwidth,clip]
{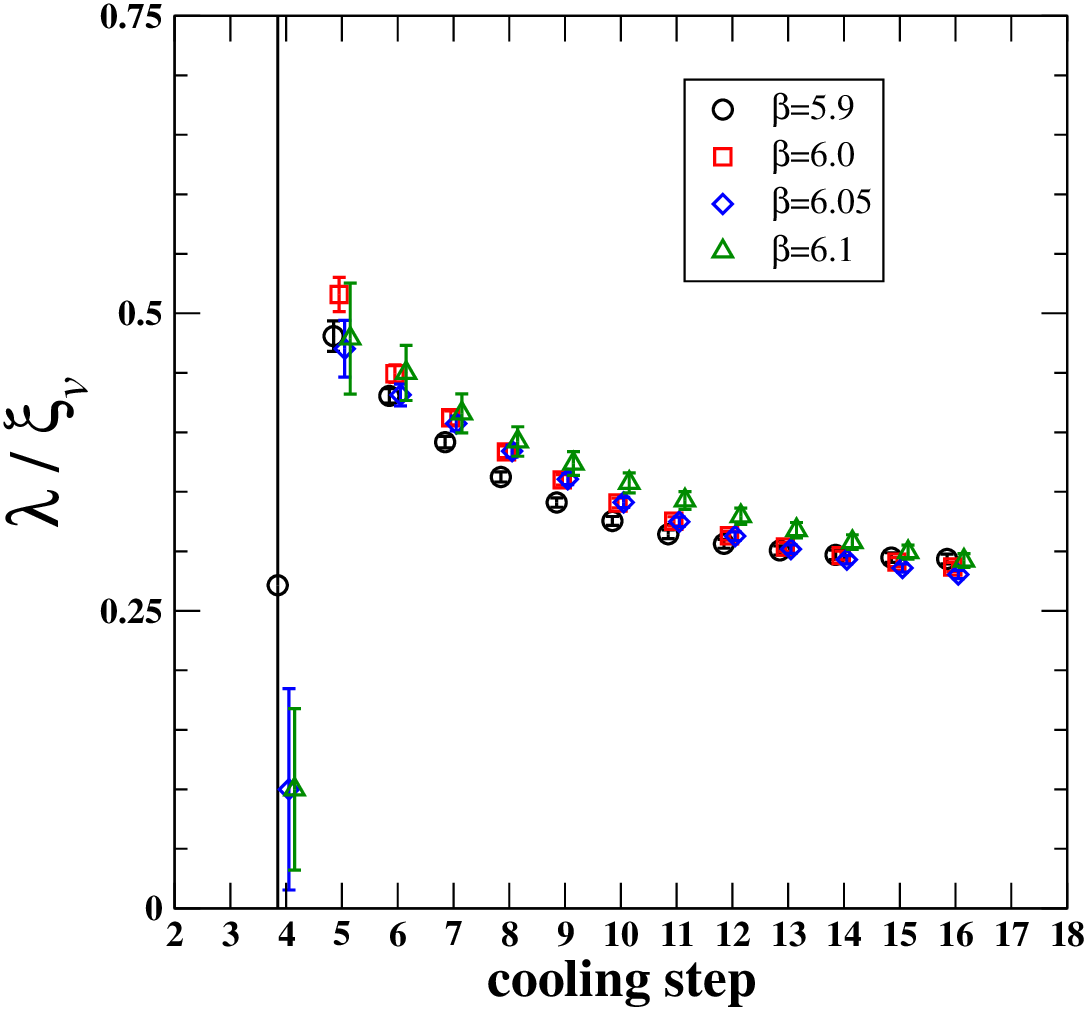} 
\caption{SU(3): $\lambda/\xi_v$ versus cooling.}
\label{lambdasucsivSU3}
\end{figure}

\begin{figure}[htb]
\includegraphics*[width=0.95\columnwidth,clip]
{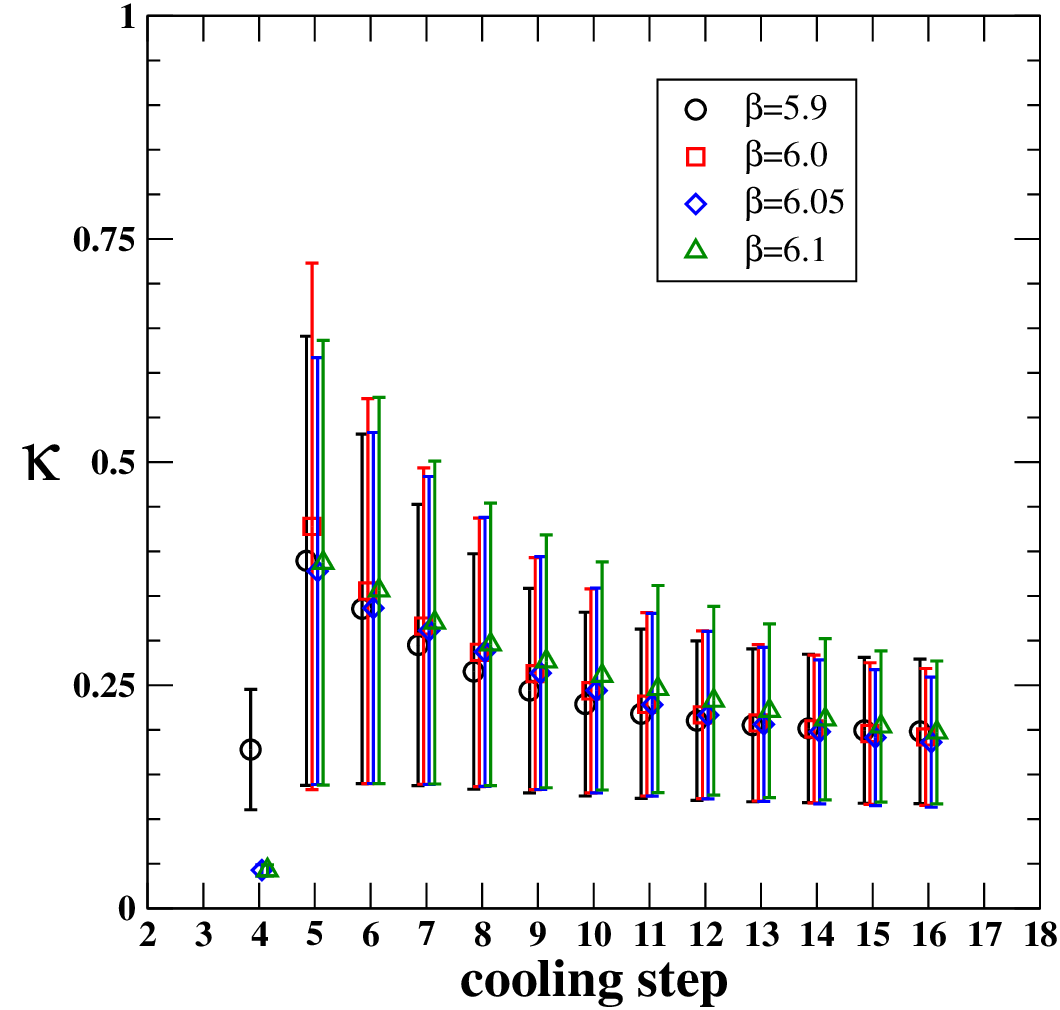} 
\caption{SU(3): $\kappa$ versus cooling.}
\label{kappaSU3}
\end{figure}

In Table II  we collect the results of our fit of Eq.~(\ref{clem2}) to 
the SU(3) data at $\beta=6.0$ for cooling steps ranging from 5 up to 16.
In Figs.~\ref{phiSU3},  \ref{muSU3},  \ref{lambdasucsivSU3},  \ref{kappaSU3} 
we show the fitted parameters versus the cooling steps. \\
In fact, we see that the parameters $\mu$, $\lambda/\xi_v$, and $\kappa$, 
display a short plateau during the controlled cooling procedure as in the
SU(2) case. Moreover, Fig.~\ref{phiSU3} shows that, at variance with the 
previous case, even the overall normalization of the transverse distribution
of the longitudinal chromoelectric field $\phi$ seems to displays an 
approximate plateau after 7--9 cooling steps in agreement with the
results of Ref.~\cite{Cardaci:2010tb}.
\section{Penetration and coherence lengths}
\label{lengths}
\begin{figure}[htb]
\includegraphics*[width=0.95\columnwidth,clip]
{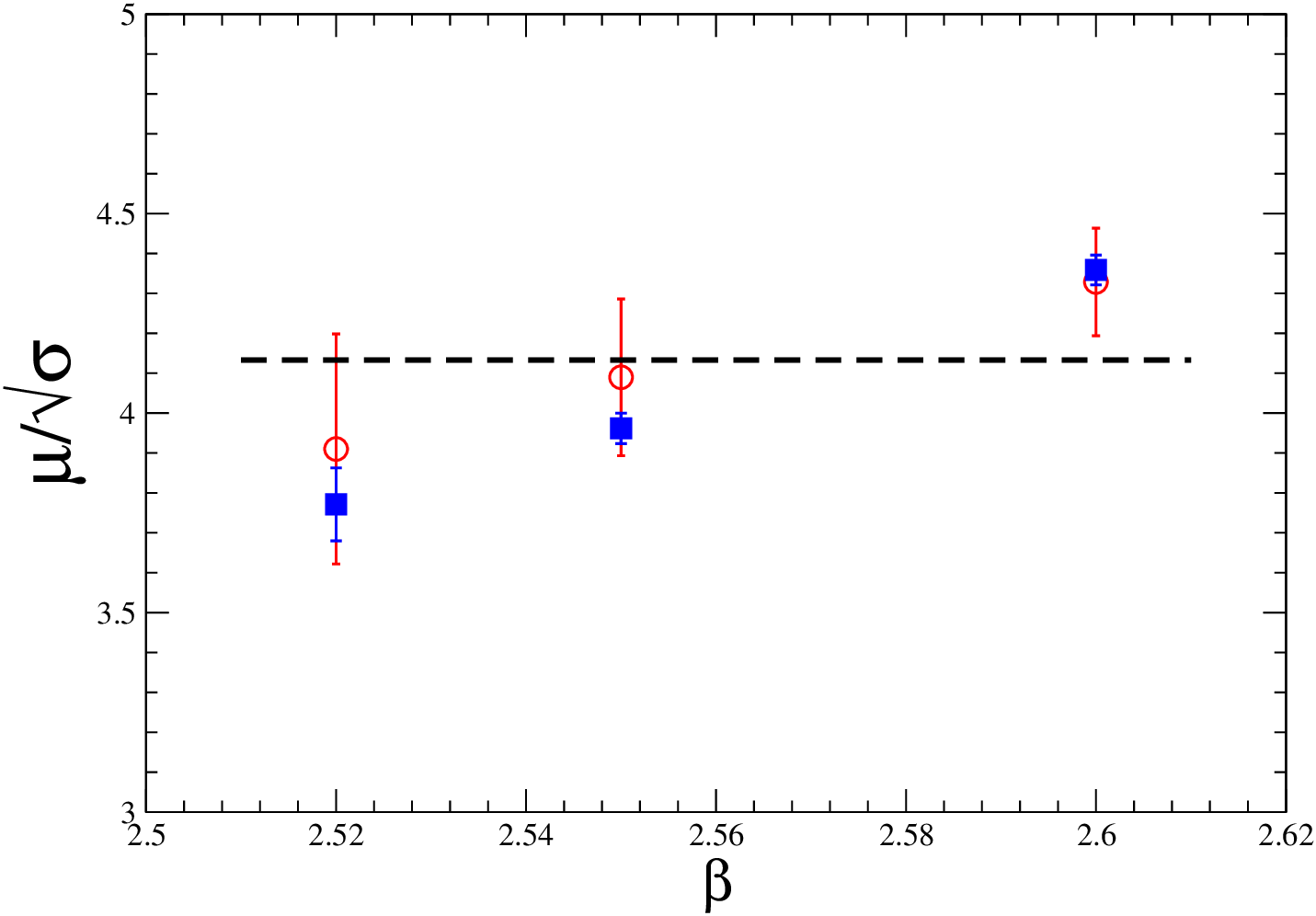} 
\caption{SU(2): $\mu/\sqrt{\sigma}$ versus $\beta$. Open circles corresponds 
to the fit with Eq.~(\ref{London}) after 8 cooling steps, full squares 
correspond to the fit with Eq.~(\ref{clem2}) after 10 cooling steps.}
\label{musigmascalingsu2}
\end{figure}
\begin{figure}[htb]
\includegraphics*[width=0.95\columnwidth,clip]
{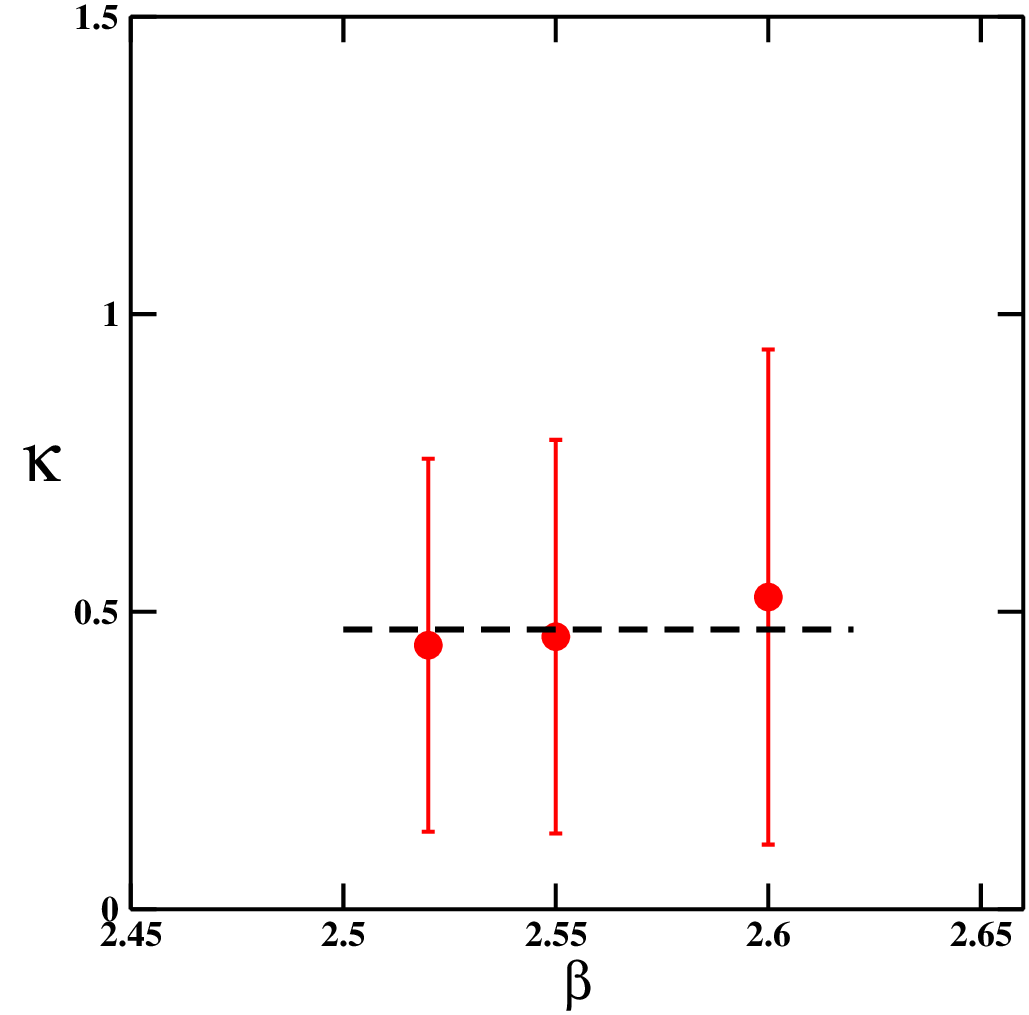} 
\caption{SU(2): $\kappa$ versus $\beta$ for 10 cooling steps.}
\label{kappascalingsu2}
\end{figure}
\begin{figure}[htb]
\includegraphics*[width=0.95\columnwidth,clip]
{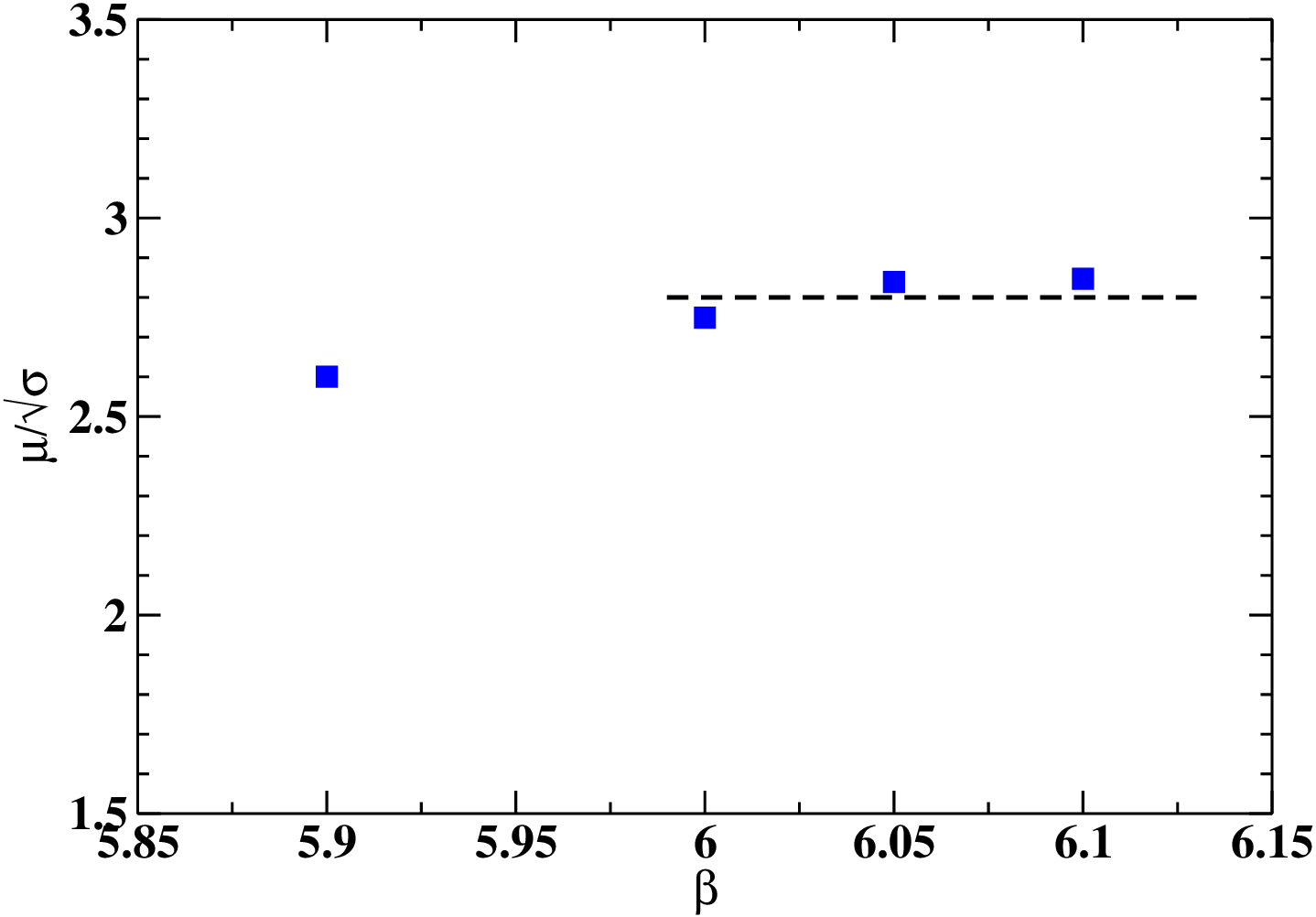} 
\caption{SU(3): $\mu/\sqrt{\sigma}$ versus $\beta$. Full squares correspond 
to the fit with Eq.~(\ref{clem2}) after 10 cooling steps.}
\label{musigmascalingsu3}
\end{figure}
\begin{figure}[htb]
\includegraphics*[width=0.95\columnwidth,clip]
{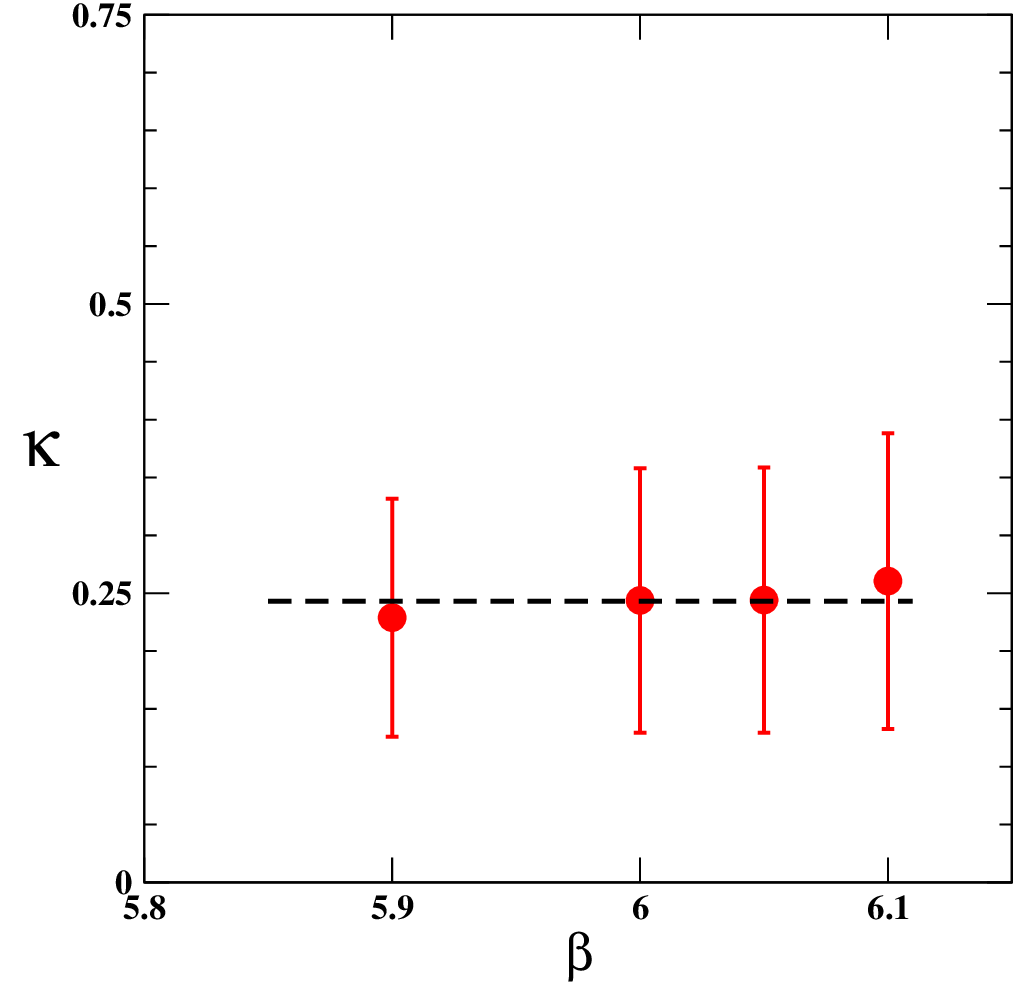} 
\caption{SU(3): $\kappa$ versus $\beta$ for 10 cooling steps.}
\label{kappascalingsu3}
\end{figure}

In Refs.~\cite{Cea:1995zt,Cardaci:2010tb} it was found that the inverse 
penetration length $\mu$ exhibits approximate scaling with the
string tension $\sigma$. To check the scaling of our new determination
of $\mu$ with the string tension, we use a parameterization for
the SU(2) string tension obtained by means of a Chebyshev polynomial 
interpolation to the string tension data collected in Table~10 of 
Ref.~\cite{Teper:1998kw}. \\
In Fig.~\ref{musigmascalingsu2} we display our determination of the ratio
$\mu/\sqrt{\sigma}$ for three different values of $\beta$. For comparison we 
also report $\mu/\sqrt{\sigma}$, where the inverse of the penetration length 
$\mu$ is obtained by fitting  the transverse profile of the longitudinal 
chromoelectric field to Eq.~(\ref{London}) after 8 cooling steps 
(for details, see Ref.~\cite{Cardaci:2010tb}). We see that our new 
determination of $\mu/\sqrt{\sigma}$ is in satisfying agreement with the 
results of Ref.~\cite{Cardaci:2010tb}.
Thus, we confirm  that $\mu$ displays an approximate scaling with the string 
tension $\sigma$. Fitting our data for $\mu/\sqrt{\sigma}$ with a constant, 
we estimate
\begin{equation}
\label{3.1}
\mu/\sqrt{\sigma}=4.133(98) \; ,
\end{equation}
where the quoted error take care also of the systematic errors due to the 
scaling violations displayed by our data. Assuming $\sqrt{\sigma}=420$~MeV,
Eq.~(\ref{3.1}) gives for the penetration length
\begin{equation}
\label{3.2}
\lambda \; = \; \frac{1}{\mu} \; = \; 0.1135(27)   \; \; {\rm fm}  \; .
\end{equation}
Moreover, we have also checked the scaling of the Ginzburg-Landau parameter  
$\kappa$ obtained through Eq.~(\ref{landaukappa}). In fact,  
Fig.~\ref{kappascalingsu2} shows that $\kappa$ is almost insensitive to 
$\beta$. By fitting the data with a constant we get
\begin{equation}
\label{3.3}
\kappa \; = \;  0.467 \; \pm \; 0.310  \; .
\end{equation}
We would like to stress that our continuum extrapolation for the penetration 
length and the Ginzburg-Landau parameter are in reasonable agreement with the 
results obtained in Refs.~\cite{Suzuki:2007jp,Schlichter:1997hw,Bali:1997cp}.
In particular, we may confirm that the Ginzburg-Landau parameter is consistent 
with the critical value $\kappa_c = \frac{1}{\sqrt{2}}$,
i.e. the SU(2) vacuum behaves as a dual superconductor which lies at the 
borderline between the type I - type II superconductor regions. \\
Also for the  SU(3) gauge theory  we studied the scaling of the ``plateau'' 
values of $\mu$ with the string tension. For this purpose, we have expressed 
our values of $\mu$ in units of $\sqrt\sigma$, using the parameterization
\bea
\label{sqrt-sigma-SU3}
\sqrt{\sigma}(g)&=&f_{{\rm{SU(3)}}}(g^2)[1+0.2731\,\hat{a}^2(g) \\
&-&0.01545\,\hat{a}^4(g) +0.01975\,\hat{a}^6(g)]/0.01364 \;, \nonumber
\eea
\[
\hat{a}(g) = \frac{f_{{\rm{SU(3)}}}(g^2)}{f_{{\rm{SU(3)}}}(g^2(\beta=6))} \;, \;
\beta=\frac{6}{g^2} \,, \;\;\; 5.6 \leq \beta \leq 6.5\;,
\]
\beq
\label{fsun}
f_{{\rm{SU(3)}}}(g^2) = \left( {b_0 g^2}\right)^{- b_1/2b_0^2} 
\, \exp \left( - \frac{1}{2 b_0 g^2}\right) \,,
\eeq
\[
b_0=\frac{11}{(4\pi)^2}\;, \;\; b_1=\frac{102}{(4\pi)^4}\;,
\]
given in Ref.~\cite{Edwards:1998xf}.
Figure~\ref{musigmascalingsu3} suggests that the ratio $\mu/\sqrt\sigma$ 
displays a nice plateau in $\beta$, as soon as $\beta$ is larger than 6. 
Accordingly, fitting the ratio $\mu/\sqrt\sigma$ to a constant we get:
\begin{equation}
\label{mu_sqrt-sigma-SU3}
\frac{\mu}{\sqrt{\sigma}} = 2.799 (38) \;,
\end{equation}
which, assuming again the standard value for the string tension 
$\sqrt{\sigma}=420$~MeV, corresponds to:
\begin{equation}
\label{3.4}
\lambda \; = \; \frac{1}{\mu} \;  =  \; 0.1676 ( 23)  \;  \; \; {\rm fm}  \; .
\end{equation}
The quoted error takes into account our estimation of systematic effects
due to the small scaling violations present in our lattice data.  
It is interesting to compare the present determination of the SU(3)  
penetration length with the one obtained previously~\cite{Cardaci:2010tb}
by fitting the lattice data to Eq.~(\ref{London}). In fact in 
Ref.~\cite{Cardaci:2010tb} we obtained $\frac{\mu}{\sqrt{\sigma}} = 2.325(5)$ 
which, unlike to the SU(2) gauge theory, seems not to agree with  
Eq.~(\ref{mu_sqrt-sigma-SU3}). This discrepancy must be ascribed to the
fact that, as discussed below, in the present case the coherence length 
exceeds the penetration length. In other words, the SU(3) vacuum behaves like 
a type I superconductor and, as we already discussed, the London equation 
Eq.~(\ref{London}) is not good enough to describe the transverse distribution 
of the longitudinal chromoelectric field. \\
As concerns the Ginzburg-Landau parameter, in Fig.~\ref{kappascalingsu3} 
we present our lattice data for different values of $\beta$. Indeed, $\kappa$
is almost insensitive to $\beta$. By fitting the data with a constant we get
\begin{equation}
\label{3.5}
\kappa \; = \;  0.243 \; \pm \; 0.088  \; ,
\end{equation}
which confirms that $\kappa < \kappa_c$ (type I superconductor). 
It is worthwhile to note that our Eqs.~(\ref{3.4}), (\ref{3.5}) are not in 
agreement with the recent determinations in Ref.~\cite{Cardoso:2010kw}, where 
it is reported $\lambda = 0.2013 (174 )$ fm and $\kappa = 1.218 (109)$.
We believe that the origin of the discrepancies resides in the use of 
different lattice operators to extract the longitudinal chromoelectric field. 
In fact, the authors of Ref.~\cite{Cardoso:2010kw} use a lattice operator 
which is sensitive to the square of the chromoelectric field instead of our 
correlation function, Eq.~(\ref{rhoW}), which measures the chromoelectric 
field strength. In any case, we believe that these discrepancies deserve 
further studies.
To summarize, we have found that the transverse behavior of the longitudinal 
chromoelectric field can be fitted according to Eq.~(\ref{clem1}) for both 
SU(2) and SU(3) gauge theories.
This allows us determine the coherence and penetration lengths. In 
Ref.~\cite{Cardaci:2010tb} it was stressed that the ratio between the 
penetration lengths respectively for SU(2) and SU(3) gauge theories recalls 
the analogous behavior seen in a different study of SU(2) and SU(3) 
vacuum in a constant external chromomagnetic background 
field~\cite{Cea:2005td}. 
In fact, in Ref.~\cite{Cea:2005td} numerical evidence was presented that the 
deconfinement temperature for SU(2) and SU(3) gauge systems in a constant 
Abelian chromomagnetic field decreases when the strength of the applied field 
increases. 
Moreover, as discussed in Refs.~\cite{Cea:2001an,Cea:2005td,Cea:2007yv}, above 
a critical strength $\sqrt{gH_c}$ of the chromomagnetic external background 
field the deconfined phase extends to very low temperatures. It was 
found~\cite{Cea:2005td} that the ratio between the critical field strengths 
for SU(2) and SU(3) gauge theories was
\begin{equation}
\label{3.6}
\frac{\sqrt{gH_c}|_{\text{SU(2)}}}{\sqrt{gH_c}|_{\text{SU(3)}}} = 2.03(17) \; .
\end{equation}
It is interesting to compare the ratio between the critical field strengths 
Eq.~(\ref{3.6}) with the analogous ratio between penetration and coherence 
lengths. Combining Eqs.~(\ref{G-L-kappa}), (\ref{3.1}), (\ref{3.3}), (\ref{mu_sqrt-sigma-SU3}), 
and~(\ref{3.5}), we readily obtain
\begin{equation}
\label{3.7}
\frac{\lambda_{\text{SU(3)}}}{\lambda_{\text{SU(2)}}} \; = \;  
\frac{\mu_{\text{SU(2)}}}{\mu_{\text{SU(3)}}} \; = 1.48 (4)  \; ,
\end{equation}
\begin{equation}
\label{3.8}
\frac{\xi_{\text{SU(3)}}}{\xi_{\text{SU(2)}}} \;  =  \;  2.84 \; \pm \; 2.15   \; .
\end{equation}
It is remarkable that the ratio between the penetration lengths, 
respectively for SU(3) and SU(2) gauge theories, agrees with the analogous 
ratio between the coherence lengths, albeit within the rather large
statistical uncertainty.  Moreover, both ratios are in fair agreement with 
the ratio between the critical field strengths, Eq.~(\ref{3.6}). \\
As stressed in the Conclusions of Ref.~\cite{Cea:2005td}, the peculiar 
dependence of the deconfinement temperature on the strength of the Abelian 
chromomagnetic field $gH$ could be naturally explained if the vacuum behaved 
as a disordered chromomagnetic condensate which confines color charges due 
both to the presence of a mass gap and the absence of color long range order, 
such as in the Feynman picture for Yang-Mills theory in (2+1) 
dimensions~\cite{Feynman:1981ss}.
The circumstance that  the ratio between the SU(2) and SU(3) penetration 
and coherence lengths agrees within errors with the above discussed ratio of 
the critical chromomagnetic fields, suggests us that the Feynman picture of 
the Yang-Mills vacuum could be a useful guide to understand the dynamics of 
color confinement. 

\section{Chromoelectric flux tube and  string tension}
\label{string-tension}

\begin{figure}[htb]
\includegraphics*[width=0.95\columnwidth,clip]
{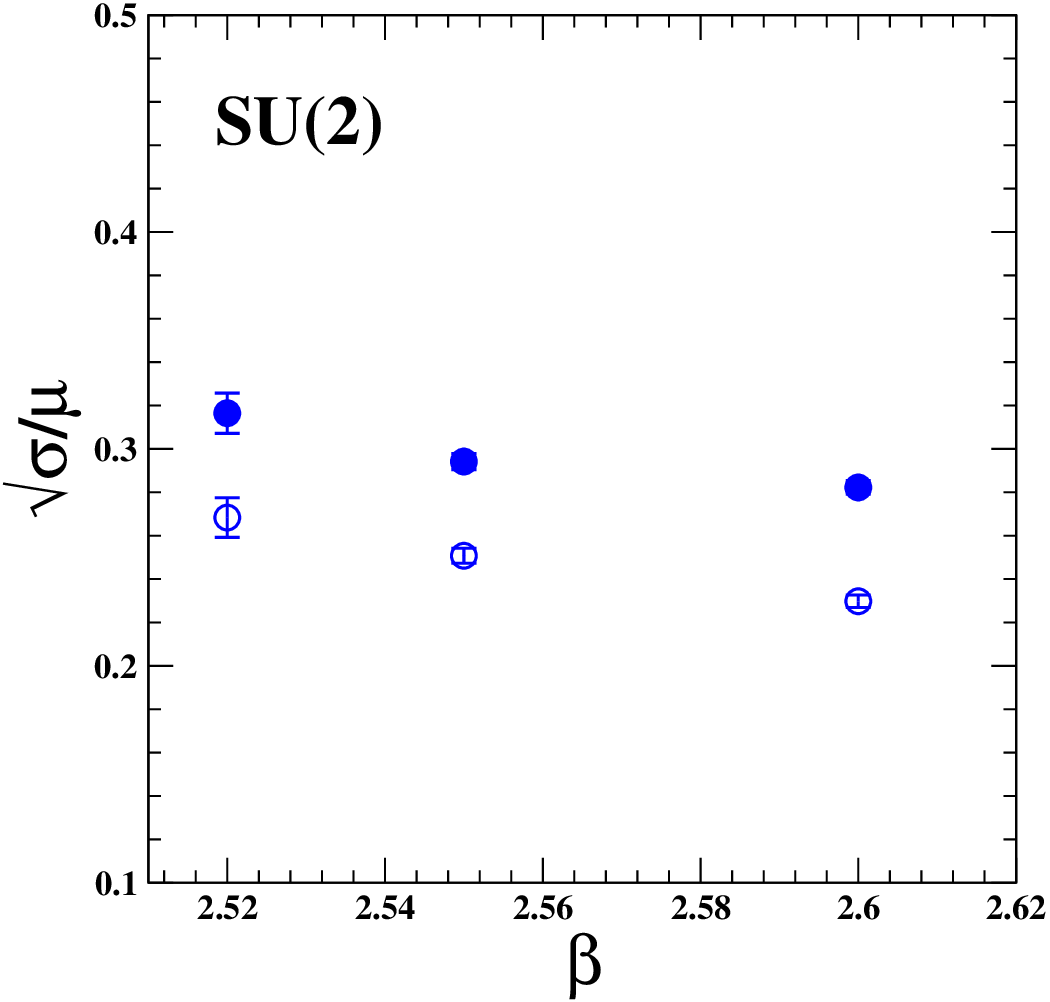} 
\caption{SU(2): $\sqrt{\sigma}/\mu$ versus $\beta$. Full points correspond to Eq.~(\ref{4.3}); open
points refer to the lattice string tension (see the discussion in the text).}
\label{sigmamuscalingsu2}
\end{figure}

\begin{figure}[htb]
\includegraphics*[width=0.95\columnwidth,clip]
{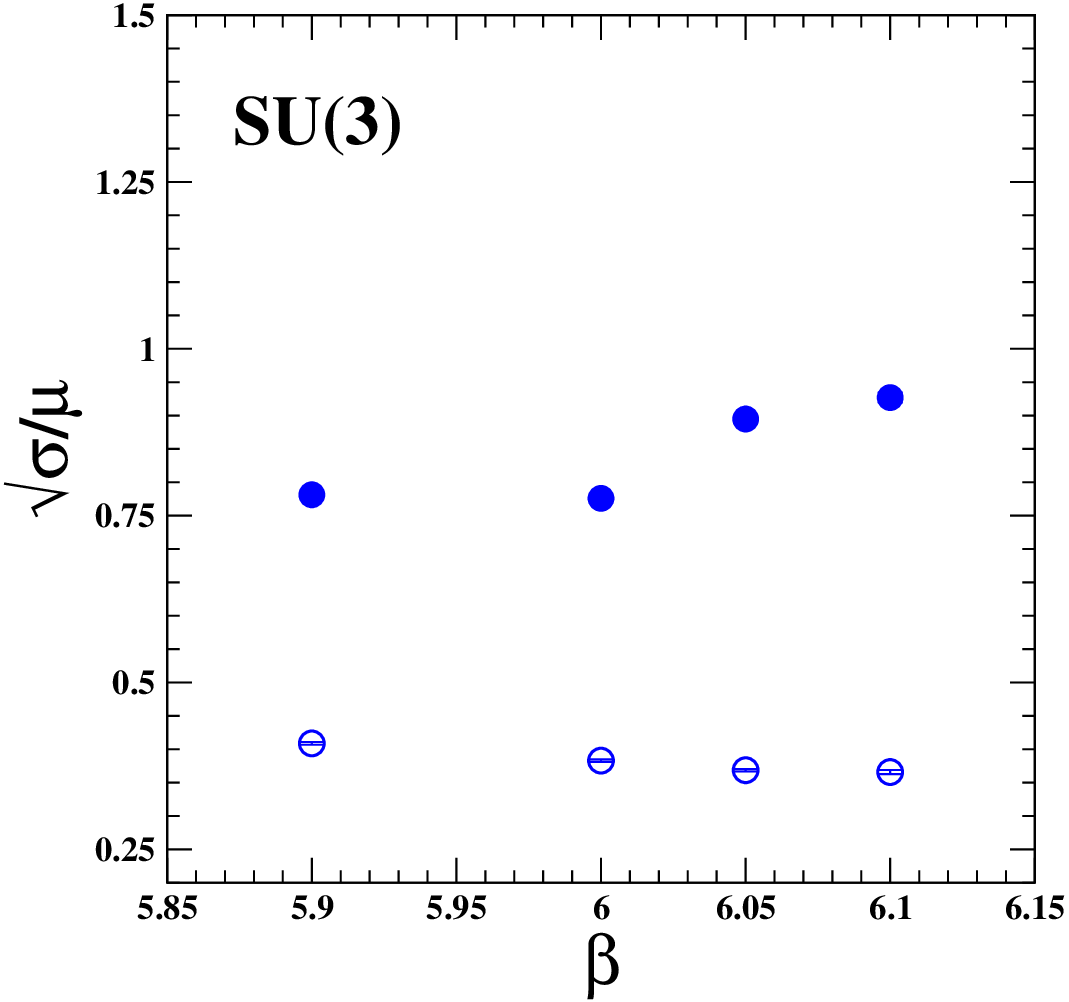} 
\caption{SU(3): $\sqrt{\sigma}/\mu$ versus $\beta$. Full points correspond to Eq.~(\ref{4.3}); open
points refer to the lattice string tension (see the discussion in the text).}
\label{sigmamuscalingsu3}
\end{figure}

We have shown~\cite{Cea:1995zt,Cardaci:2010tb} that the color fields of a 
static quark-antiquark pair are almost completely described by the 
longitudinal chromoelectric field, which in turn is approximately constant 
along the flux tube.  This means that the long-distance potential acting on
the color charges is linear. 
Using our data and the parameterization Eq.~(\ref{clem2}) for the 
chromoelectric flux tube, we are able to compute the string tension given as 
the energy stored into the flux tube per unit length:
\begin{equation}
\label{4.1}
\sigma_{E_l}   \simeq \frac{1}{2} \; \int d^2 x_t E^2_l(x_t) \,,
\end{equation}
where, to avoid confusion, we have denoted  the flux-tube string tension  as 
$\sigma_{E_l}$, while $\sigma$ will indicate the lattice string tension. 
It is worth to note that the string tension $\sigma_{E_l} $ defined by 
Eq.~(\ref{4.1}) does not depend on $x_l$ as long as the longitudinal 
chromoelectric field is constant along the flux tube. Obviously this last 
condition is not strictly fulfilled on a finite lattice. From 
Eqs.~(\ref{4.1}) and~(\ref{clem2}) we obtain an explicit relation between the 
string tension and the parameters $\phi$, $\mu$, and $\alpha$ of the fit 
Eq.~(\ref{clem2}) to the chromoelectric flux tube profile:
\begin{equation}
\label{4.2}
\sigma_{E_l}  = 
 \frac{1}{4 \pi}  \frac{\phi^2 \mu^4}{\alpha^2}  \frac{1}{K_1^2(\alpha)} 
\int_0^\infty dr \, r \,  K_0^2((\mu^2 r^2 + \alpha^2)^{1/2}) \; .
\end{equation}
After performing the integration, we get:
\begin{equation}
\label{4.3}
\frac{\sqrt{\sigma_{E_l} }}{\mu} = \left( \frac{\phi^2}{8 \pi} 
\left(1 - \frac{K_0^2(\alpha)}{K_1^2(\alpha)}\right) \right)^{1/2} \; .
\end{equation}
Naively, one expects that the string tension defined as the energy 
per unit length stored into the flux tube chromoelectric field Eq.~(\ref{4.3}) 
should agree, at least approximatively, with the string tension measured on 
the lattice. To check this, in Figs.~(\ref{sigmamuscalingsu2}) 
and~(\ref{sigmamuscalingsu3})  we compare the flux-tube string tension in
Eq.~(\ref{4.3}) with the lattice string tension (obtained as detailed in the
previous Section) for SU(2) and SU(3) gauge theories, respectively.
It  is evident from Figs.~(\ref{sigmamuscalingsu2}) and~(\ref{sigmamuscalingsu3}) 
that
\begin{equation}
\label{4.4}
\sigma_{E_l} \; > \;  \sigma   \; 
\end{equation}
for both SU(2) and SU(3) gauge theories. At first sight this result looks 
quite surprising. In fact, the lattice string tension should contain the total 
energy per unit length stored into the flux tube. As a consequence we can 
write
\begin{equation}
\label{4.5}
\sigma  \;  \simeq  \; \;  \sigma_{E_l}  \; + \;  \sigma_{\rm cond}   \; ,
\end{equation}
where $\sigma_{\rm cond}$ takes into account the contribution due to the order 
parameter condensate.
We may, in turn, obtain an estimate of this contribution to the 
total string tension as follows. Since within the vortex core the order 
parameter condensate vanishes, we have
\begin{equation}
\label{4.6}
\sigma_{\rm cond} \;  \simeq \;  - \;  \pi \, \xi^2\, \varepsilon_{\rm cond}\; 
\end{equation}
where $\varepsilon_{\rm cond}$ is the condensation energy density and $\xi$ is 
approximately the vortex core size. Note that the minus sign is due to the 
loss  of condensation energy in the normal region with respect to the
confining vacuum where the order parameter is nonzero. Now, it is 
usually assumed that in the confining vacuum it is energetically favored to 
have a condensation of the order parameter as in ordinary BCS
superconductors where the superconducting transition is energetically driven 
by the coherent condensation of Cooper pairs.
Therefore, one is led naturally to suppose that $\varepsilon_{\rm cond} < 0$. 
Thus we see from Eqs.~(\ref{4.5}) and (\ref{4.6}) that one should obtain  
$\sigma_{E_l} <  \sigma$. On the contrary our numerical results clearly 
indicate that $\sigma_{E_l}$ exceeds $\sigma$. The only possible 
conclusion we 
can derive is that the order parameter condensation energy is positive.
Thus, the confining transition must be driven by disordering the gauge system. 
In other words, even though the condensation of the confining order parameter 
costs energy there is a huge number of degenerate physical configurations such 
that the configurational entropy easily overcomes the energy cost. This 
means that the deconfining transition is an order-disorder transition, much like 
the Berezinskii-Kosterlitz-Thouless transition than the BCS superconducting 
transition. 
It is remarkable that this conclusion reinforces our previous picture of the 
confining vacuum that behaves like a disordered chromomagnetic condensate 
which confines color charges due both to the presence of a mass gap and the 
absence of color long range order, such as in the Feynman qualitative 
picture~\cite{Feynman:1981ss}.
\section{Conclusions}
\label{conclusions}
In the present paper we studied the chromoelectric field distribution
between a static quark-antiquark pair in SU(2) and SU(3) pure gauge theories.
By means of the connected correlator given in Eq.~(\ref{rhoW}) we were
able to compute the chromoelectric field that fills the flux tube along the 
line joining a quark-antiquark pair. 
Since our connected correlator is sensitive to the field strengths instead of 
the squared field strength, we were able to follow the transverse shape of 
the color fields up to sizable distances.
Using some dated results in ordinary superconductivity based on a simple 
variational model for the magnitude of the normalized order
parameter of an isolated vortex, we proposed that the transverse behavior of 
the longitudinal chromoelectric field can be fitted according to 
Eq.~(\ref{clem2}), which allowed us to get informations on the penetration
and coherence lengths. In fact we found that our Eq.~(\ref{clem2}) is able to 
reproduce the transverse distribution of the longitudinal chromoelectric field 
in the whole available region. In the case of the SU(2) gauge theory we argued 
that the confining vacuum behaves as a dual superconductor which lies at the 
borderline between the superconductor  type I - type II regions. On the other 
hand, we found that the SU(3) vacuum belongs to the superconductor type I 
region. We found that the ratio between the penetration lengths respectively 
for SU(3) and SU(2) gauge theories agrees with the analogous ratio between 
the coherence lengths, albeit within the rather large statistical uncertainty, 
and both ratios are in fair agreement with the ratio between the critical 
chromomagnetic fields.
Finally, we suggested that the deconfining transition resembles the 
order-disorder Berezinskii-Kosterlitz-Thouless transition and that the 
confining vacuum behaves like a disordered chromomagnetic condensate in  
agreement with the Feynman qualitative picture of the Yang-Mills vacuum.
%

%

\end{document}